\documentclass[12pt]{iopart}
\usepackage{epsfig}
\def\bea{\begin{eqnarray}} 
\def\beann{\begin{eqnarray*}}
\def\beq{\begin{equation}}
\def\eea{\end{eqnarray}}
\def\eeann{\end{eqnarray*}}
\def\eeq{\end{equation}}
\def\nn{\nonumber}

\def\eps{\epsilon}
\def\sig{\sigma}

\def\bi{\bibitem}

\begin{document}

\title{Can quark
%-gluon exchange 
effects be observed in intermediate heavy ion collisions?}
\author{ D.T. da Silva and D. Hadjimichef}
\address{ Instituto de F\'{\i}sica e Matem\'{a}tica,Universidade Federal de Pelotas, 
Pelotas, R.S. Brazil}

\begin{abstract}
In recent years a tentative description of the short-range
part of hadron interactions with constituent
quark interchange has been developed providing an alternative approach to meson physics. 
Quark interchange plays a role, for example, in the nucleon-nucleon ($NN$) phase-shifts
and cross-section.
In heavy ion collision simulations at intermediate energies  one of the main features 
is the $NN$ cross-section in the collisional term, where in most
cases it is an input adjusted to the free space value.
In this paper we introduce the quark degrees of freedom to the $NN$  cross-section in the  
Vlasov-Uehling-Uhlenbeck (VUU) model and  explore the possibility that these 
effects appear in the observables at lower energies.
\end{abstract}

\submitto{\JPG}

\maketitle

\section{Introduction}

For a long time low energy hadron-hadron interactions, in particular the $NN$
interaction, have been studied in terms of meson exchange models
employing baryon and meson degrees of freedom.  Such models have
achieved a good description of the empirical data.
A common undesirable feature in all meson exchange models is the
necessity of a (semi)phenomenological parametrization of the short-range
part of the interaction. However, since the advent of quantum
chromodynamics (QCD) there is the hope that the short-range part of the 
hadron-hadron interaction can
be derived from first principles. However, because of complications
due to the nonperturbative nature of the QCD interactions at low energies,
the use of models is a practical necessity for making progress in this
direction.
 
There is a great variety of quark models that can describe with
reasonable success single-hadron properties. Therefore, there arises the
natural question of to what extent a model which gives a good hadron
structure is, at the same time, able to describe the hadron-hadron
interaction and in a larger context, intermediate energy nucleus-nucleus collisions.  
In this respect, in the recent past several studies of the $NN$ interaction have employed the
non-relativistic quark model, {\it a la} Isgur and Karl, which is one of the
most successful quark models for hadron structure. The main ingredient
of such calculations is the one-gluon exchange interaction.
The results are encouraging as far as the very short-range part is
concerned which is reflected in the successful description of $NN$
scattering S-wave phase shifts. For the intermediate-range part 
progress on the spin-orbit problem in the $NN$ system was made by including, in addition
to the one-gluon-exchange interaction, a scalar-isoscalar meson coupled to
quarks in a chirally symmetric way \cite{Tue}. 
More recently, calculations in the $KN$ 
system became an important approach, due to the fact that 
one-pion exchange is absent  and contributions from $2\pi$ exchange seem to 
be weaker than in the $NN$ system. The quark-gluon exchange in the $KN$ system has been 
studied by Barnes and Swanson \cite{kn1}, Silvestre-Brac and collaborators \cite{kn2}-\cite{kn2d} and
Hadjimichef, Haidenbauer and Krein \cite{kn3}.
In summary, these studies lead to some general questions such as: where do we place the frontier between 
quark-gluon physics and effective theories?
To what extent the underlying quark-gluon physics,
studied at lower energies in the context of  hadron interactions,
can play a role in the  complex dynamics related to heavy ion collisions?
  Can we  have a description in which one could
smoothly pass from high  energy physics to an effective low energy representation?
Of course to answer these questions would represent a major step in hadron physics. In a 
more limited context it would be interesting  to extend the  quark-gluon exchange  
approach to a many-body $NN$ system such as a nucleus and study 
in heavy ion collisions.

Little is known about hadronic matter at
finite temperatures and densities other than the nuclear ground state
density $\rho _{0}=0.16$ fm$^{-3}$. 
 Unfortunately, there exists, at the moment, no theoretical
model that consistently provides an understanding of the reaction dynamics of heavy ion collisions over
the entire energy range.
It is believed that the quark-gluon plasma
can be observed in high energy collisions, where temperature and
density are large enough to deconfine quarks and gluons. 
At lower energies heavy ion collisions have be used to investigate 
the appearance of collective effects and to probe the
conditions related to the nuclear equation of state \cite{stocker1,stock}.
Theoretically these studies have been performed by numerical simulations in many models.
For example, transverse and longitudinal
collective flow as well as azimuthal distributions provides complementary information which can
be used to evaluate the details of microscopic models such as Boltzmann equation 
\cite{boltzmann,boltzmann2},
Vlasov-Uehling-Uhlenbeck \cite{stocker2,VUU,VUU2} and quantum molecular dynamics \cite{stocker3}.
One of the main ingredients in these models is the nucleon-nucleon cross-section, where it turns
out that  many observables are very sensitive to its value. It is usual to assume as an input
to simulations the free $NN$ cross-section. However recent studies on 
collective flow have indicated a density dependent in-medium reduction from its value in free space  
\cite{in-medium,in-medium1}.
 
 In this paper we shall extend the study of Ref. \cite{hadrons} and
 develop this aspect considering quark-gluon exchange corrections to the 
nucleon-nucleon cross-section in the VUU equation as a first approach.
The procedure we shall use, in order to introduce these corrections,  is known as the
  Fock-Tani formalism  (FTf), which is a
method that employs a second quantization formalism to problems
 where the internal degrees of freedom of composite particles cannot validly
be neglected.

\section{The model}

The Vlasov-Uehling-Uhlenbeck (VUU) equation is a differential equation for
the classical one-body phase-space distribution function $f({\bf r},{\bf p},t)$ 
corresponding to the classical limit of the Wigner function. The VUU
equation takes a familiar form
\begin{eqnarray}
\fl
\frac{\partial f}{\partial t}+{\bf v}\cdot \bigtriangledown
_{r}f-\bigtriangledown _{r}U\cdot \bigtriangledown _{p}f &=&-\int \frac{%
d^{3}p_{2}\,d^{3}p_{1}^{\prime }\,d^{3}p_{2}^{\prime }}{(2\pi )^{6}}\;\sigma
\;v_{12}\,\,\delta ^{3}(p+p_{2}-p_{1}^{\prime }-p_{2}^{\prime })
\nonumber
\\
&&\times \left[ f\,\,f_{2}(1-f_{1}^{\,\prime })(1-f_{2}^{\,\,\prime
})-f_{1}^{\,\prime }\,\,f_{2}^{\,\,\prime }(1-f)(1-f_{2})\right] 
\label{vuu}
\end{eqnarray}
where $f_{2}^{\;\prime }$ is a stands for  $f({\bf r},{\bf p}
_{2}^{\prime },t)$ and correspondingly for the other terms.The two-body interactions are 
divided in a short-range and long-range part. The short-range interactions are 
regarded as corresponding to the hard-core binary
collisions. The long-range interactions are assumed to be given by a potential
with origin in a density-dependent mean field, formed by the nuclear matter, around the particle 
in use. The potential $U$ may be written in terms of the modified Skyrme forces \cite{stocker1}
\bea 
U=\alpha \frac{\rho}{\rho_0}   +\beta\left(\frac{\rho}{\rho_0} \right)^{\gamma}\,\,.
\label{skyrme}
\eea
The conditions used to fix the 
three parameters $\alpha$, $\beta$ and $\gamma$ are: $(i)$ the 
ground state energy must assume the correct value for nuclear matter (-16 MeV); $(ii)$ the ground
state at $\rho=\rho_0$ has to be a minimum; $(iii)$ the compression modulus $\kappa$ should be
some hundred MeV. The magnitude of $\kappa$ can be obtained from the measure of the radius of 
the giant monopole resonances in nuclei.

In order to include the quark corrections to the VUU equation we 
shall briefly outline the main features of the Fock-Tani approach to baryon-baryon
interactions.
One starts with the Fock space representation of the system, using creation
and annihilation operators of the elementary hadron constituents particles.
A one-baryon state,
with momentum ${\bf P}$, internal energy $\epsilon$, spin projection $M_S$
and isospin projection $M_T$, can be written as 
\begin{equation}
|\alpha \rangle \equiv |{\bf P}, \epsilon, M_S, M_T\rangle
=B_{\alpha}^{\dag}|0\rangle\;,  \label{1}
\end{equation}
where $|0\rangle$ is the vacuum (no quarks) state and $B_{\alpha}^{\dag}$ is
the baryon creation operator 
\begin{equation}
B_{\alpha}^{\dag} = \frac{1}{\sqrt{3!}} \Psi_{\alpha}^{\mu_1\mu_2\mu_3}q_{%
\mu_1}^{\dag}q_{\mu_2}^{\dag} q_{\mu_3}^{\dag}\;.  \label{2}
\end{equation}
We use the convention that a sum over repeated indices is implied. The
indices $\mu_i$ denote the spatial, spin-flavor, and color coordinates of
the i-th quark. The baryon bound-state wave functions are taken
orthonormalized.
Using the quark relations and orthonormality condition for $\Psi$ above, one
can show the following anti-commutation relations for the baryon operators: 
\begin{eqnarray}
&&\{B_{\alpha},B_{\beta}\}=\{B^{\dag}_{\alpha},B^{\dag}_{\beta}\}=0 
\nonumber \\
&&\{B_{\alpha},B_{\beta}^{\dag}\}=\delta_{\alpha\beta}-\Delta_{\alpha\beta}%
\;,  \label{6}
\end{eqnarray}
where 
\begin{equation}
\Delta_{\alpha\beta}=3\Psi_{\alpha}^{*\mu_1\mu_2\mu_3}\Psi_{\beta}^
{\mu_1\mu_2\nu_3}q_{\nu_3}^{\dag}q_{\mu_3}-\frac{3}{2}\Psi_{\alpha}^
{*\mu_1\mu_2\mu_3}\Psi_{\beta}^{\mu_1\nu_2\nu_3}q_{\nu_3}^{\dag}q_{\nu_2}^
{\dag}q_{\mu_2}q_{\mu_3}\,\,\,.  \label{8}
\end{equation}

Observe that the baryon operators $B_{\alpha}$ do not satisfy the canonical
anti-commutation relations for fermions. The extra term $\Delta_{\alpha\beta}$
 appearing in the rhs of Eq.~(\ref{6}) results from the
composite nature of the baryons. The presence of this term complicates the
application of the usual field theoretic techniques to the 
$B_{\alpha}$ and $B_{\alpha}^{\dag}$ operators, making these operators
inconvenient dynamical variables. The change of representation
of the FTf  is implemented by means of a
unitary transformation ${\cal U}$, such that the composite particle operators $%
B_{\alpha}$ and $B^{\dag}_{\alpha}$ are re-described by ``ideal" baryon
operators $b_{\alpha}$ and $b^{\dag}_{\alpha}$. By
definition, the ideal baryon operators satisfy canonical anti-commutation
relations: 
\begin{eqnarray}
&&\{b_{\alpha}, b_{\beta}\}=\{b^{\dag}_{\alpha},b^{\dag}_{\beta}\}=0 
\nonumber \\
&&\{b_{\alpha}, b^{\dagger}_{\beta}\}=\delta_{\alpha \beta}\;,  \label{bcom}
\end{eqnarray}
and are kinematically independent from the quark operators
$\{q_{\mu},b_{\alpha}\}=\{q_{\mu},b^{\dagger}_{\alpha}\}=0 $.

The transformation from the physical space to the ideal space is performed by a unitary 
operator. The operator ${\cal U}$ is given by: 
\begin{equation}
{\cal U}=\exp\left[- {{\frac{\pi }{2}} 
\left( b^{\dag}_{\alpha}O_{\alpha} - O_{\alpha}^{\dag} b_{\alpha} 
\right)
} \right]\;,  \label{uandf}
\end{equation}
where $O_{\alpha}$ is an operator given in terms of the $B_{\alpha}$ and $%
B^{\dag}_{\alpha}$ and $\Delta_{\alpha\beta}$ of Eq.~(\ref{8}) as: 
\begin{eqnarray}
O_{\alpha}= B_{\alpha} +\frac{1}{2}\Delta _{\alpha\beta}B_{\beta} -\frac{1}{2%
}B^{\dag}_{\beta}[\Delta_{\beta\gamma},B_{\alpha}]B_{\gamma}\;.  \label{ft12}
\end{eqnarray}
Given this, the evaluation of the transformed quark operators is
straightforward. Full details of the derivation of the iterative solution
is  presented elsewhere~\cite{annals}; here we simply present the final results.
The  $O_{\alpha}$ operator is built in  such a way that a {\em single} 
real-baryon state $|\alpha\rangle$ is transformed into
a {\em single} ideal-baryon state $|\alpha)$ : 
\begin{equation}
|\alpha)\equiv
{\cal U}^{-1}B^{\dag}_{\alpha}|0\rangle 
=b^{\dagger}_{\alpha}|0)\;.  \label{single}
\end{equation}

In order to discuss nucleon-nucleon scattering, one needs to specify the
general form of the microscopic quark Hamiltonian. For our purposes here,
the microscopic Hamiltonian can be written in terms of the quark operators
as 
\begin{eqnarray}
H= K(\mu)\; q_{\mu}^{\dagger }q_{\mu} +\frac{1}{2} V_{qq}(\mu \nu; \sigma\rho) \; q_{\mu}
^{\dagger }q_{\nu}^{\dagger }q_{\rho} q_{\sigma}\;,  \label{ft23}
\end{eqnarray}
where $K$ is the kinetic energy and $V_{qq}$ is the quark-quark interaction.
Note that this is a general expression for $H$,  which in this form, a variety of 
quark-model Hamiltonians  used in the literature can
be written. In our calculation we shall use, for $V_{qq}$, the 
spin-spin hyperfine component of the perturbative one gluon interaction
\bea
V_{qq} =  -\,\frac{8\pi\alpha_{s}}{3\,m_{i}m_{j}} \,
 {\bf S}_{i}\cdot{\bf S}_{j} \,\,\, {\cal F}^{\,a}_{i}\,\, {\cal F}^{\,a}_{j}
 \,\,\,,
\label{vss}
\eea
where ${\cal F}^{\,a}_{i}=\lambda^{a}_{i}/2$ are the Gell-Mann matrices.  There is a considerable 
literature related to free $NN$ scattering with quark-interchange and in many of these models the 
quark-quark potential is much more elaborated than the potential in Eq. (\ref{vss}) 
(including Coulomb, spin-orbit, tensor, confinement terms and eventually meson coupling to quarks). 
The lesson taken from all of these works is that the dominant term for the short-range repulsion is 
basically the spin-spin term from the  one gluon exchange potential. Its strong influence is seen, 
for example, in the $^{1}S_{\,0}$ partial-wave where repulsion 
increases when $\lambda$ is increased. In these models $\lambda$ is a free parameter which usually 
ranges from 0.2 fm to 0.5 fm or even 0.6 fm (see Ref. \cite{holinde} ).
This behavior also happens in the $KN$ system \cite{kn3}. In this 
perspective a very first step in order to include quark corrections to the VUU equation would be 
to use the dominant term in  quark-quark interaction to simulate the short-range hard-core and 
verify its effect on the observables.

\begin{figure*}[htb]
\epsfxsize=25pc 
\hspace{2cm}
\epsfbox{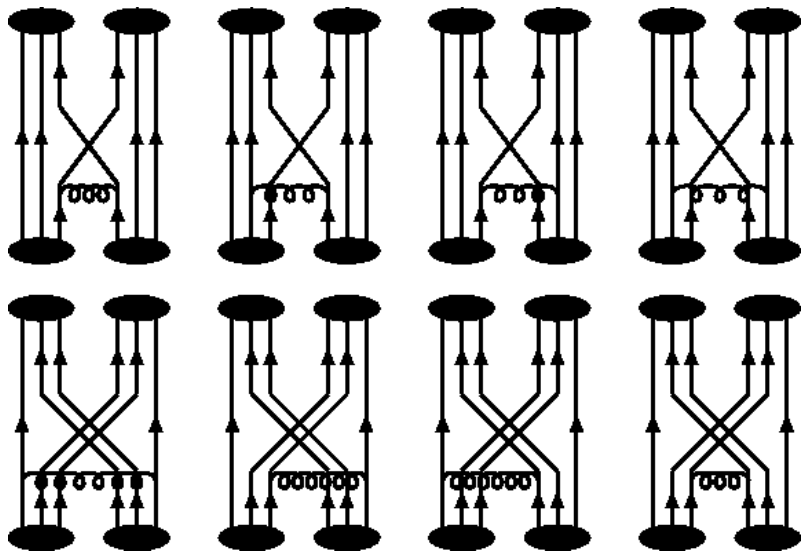}
\caption{Diagrams representing the scattering amplitude $h_{fi}$ for $NN$ 
interaction with quark interchange.}
\label{diagram}
\end{figure*}

The effective baryon-baryon Hamiltonian is obtained from the expansion retaining
the lowest order terms in $\Psi$:
\begin{eqnarray}
H_{bb}=\Psi^{*\mu\nu\lambda}_{\alpha}H(\mu\nu;\sigma\rho)
\Psi^{\sigma\rho\lambda}_{\beta}b^{\dag}_{\alpha}b_{\beta}+ \frac{1}{2}
\;V_{bb}(\alpha\beta;\gamma\delta)\;b^{\dag}_{\alpha}b^{\dag}_{\beta}
b_{\delta}b_{\gamma}\;,  \label{ft27}
\label{hbb}
\end{eqnarray}
where $V_{bb}$ is an effective baryon-baryon potential which is of $%
{\cal O}(\Psi^4)$. The scattering $T$-matrix can obtained from
Eq. (\ref{hbb})
\bea
T(\alpha\beta;\gamma\delta)=(\alpha\beta|V_{bb}|\gamma\delta)
\eea
Due to translational invariance, the $T$-matrix element is written as a momentum conservation
delta-function, times a Born-order matrix element, $h_{fi}$
\bea
T(\alpha\beta;\gamma\delta)=\delta^{(3)}( {\bf P}_{f}-{\bf P}_{i})\,h_{fi}
\eea
where ${\bf P}_{f}$ and  ${\bf P}_{i}$ 
are the final and initial momenta of the two-nucleon system. The scattering amplitude 
$h_{fi}$ is a function of the nucleon's wave-function written as
\bea
\fl
\Psi _\alpha ^{\mu _1\mu _2\mu _3}\equiv 
\frac{{\cal N }({\bf p}_\alpha )}{\sqrt{3!\,\cdot 18}}
\epsilon ^{c_{\mu _1}c_{\mu _2}c_{\mu _3}}
\,
\psi_\alpha ^{I_{\mu_1}I_{\mu _2}I_{\mu _3}}\,
\phi ({\bf p}_1)\,\phi ({\bf p}_2)\,\phi ({\bf p}_3)\,
\,\delta ({\bf p}_\alpha -{\bf p}_1-{\bf p}_2-{\bf p}_3) 
\label{qbdf5}
\eea
with $\eps^{c_{\mu _1}c_{\mu _2}c_{\mu _3}}$ the antisymmetric  color tensor; 
$\psi_\alpha ^{I_{\mu_1}I_{\mu _2}I_{\mu _3}}$ spin-isospin wave-function;
$\phi$ the quark's spatial wave-function and  ${\cal N}({\bf p}_\alpha )$ the norm. The 
function $\phi$ can be defined as a Gaussian wave-function defined by
\bea
\phi ({\bf p})&=&\left(\frac{\lambda}{\sqrt{\pi}}\right)^{3/2} \,
\exp \left( -\frac{\lambda^2 p^2}{2} \right),
\label{qbdf6}
\eea
then ${\cal N}$ is 
\bea
{\cal N}({\bf p})&=&\left(\frac{3\pi}{ \lambda^{2}}\right)^{3/4}\,
\exp \left( \frac{\lambda^2 p^2}{6}\right). 
\label{qbdf6.1}
\eea
It can be shown that the width parameter $\lambda$ is related to the $rms$ radius of the nucleon 
$\langle r^2 \rangle=\lambda^2$.
A diagrammatic representation for $h_{fi}$ can be seen in Fig. 
\ref{diagram}. The connection between the cross-section and the scattering
amplitude $h_{fi}$  is shown in Ref. \cite{annals}-\cite{QBD2}. We shall use this result in order
to evaluate  nucleon-nucleon scattering cross-section in
the Fock-Tani approach:
\bea
\sigma_{NN} =\frac{4\pi ^{5}\,s}{s-4m_{N}^{2}}
\int_{-(s-4m_{N}^{2})}^{0}\,dt\,|h_{fi}|^{2}
=\frac{4\pi ^{5}\,s}{s-4m_{N}^{2}}\;k_{ss}^{2}\;{\cal I}_{NN}
\label{cross}
\eea
where ${\kappa}_{ss}=8\pi\alpha_{s}/3m^{2}_{q}(2\pi)^{3}$ and
 the quark model the ratio $\alpha_{s}/m_{q}^{2} $ is 
related to the width of the Gaussian wave-function $\lambda$ \cite{oka1,oka2} by
\bea
\frac{\alpha_s}{m_{q}^{2}}=\frac{3\sqrt{2\pi}}{4} 
\left(m_{\Delta}-m_{N}\right)\;\;\lambda^{3}.
\label{d-n}
\eea
As a consequence  of using  a Gaussian wave-function to represent the bound-state
is that the integral ${\cal I}_{NN}$ in the cross-section (\ref{cross}) has an analytical 
expression:
\bea
\fl 
{\cal I}_{NN}&=&
\frac{3\,}{2\,\lambda^2}\left( \chi_{1}^{2} + \chi_{5}^{2} \right) 
\left[1- e^{\frac{2\,\lambda^2\,\left( 4\,m_{N}^2 - s \right) }{3}}     \right]
\nn\\
&& 
+ 
  \frac{144 }{5\,\lambda^2}
{\sqrt{\frac{3}{11}}}\,
     \left[ \chi_{5} ( \chi_{2}\,  +  \chi_{3}\,)  + 
 \chi_{1}\,( \chi_{6} +     \chi_{7}) \right]
\left[e^{\frac{8\,\lambda^2\,\left( 4\,m_{N}^2 - s \right) }{33}}
-e^{\frac{13\,\lambda^2\,\left( 4\,m_{N}^2 - s \right) }{33}}\,
           \right] 
\nn\\
&&+ 
 \frac{144 }{17\,\lambda^2} 
{\sqrt{\frac{3}{11}}}\,
     \left[  \chi_{1}\, (\chi_{2} +   \chi_{3} )
+  \chi_{5}\, (\chi_{6} +  \chi_{7}) \right]
\left[
e^{\frac{2\,\lambda^2\,\left( 4\,m_{N}^2 - s \right) }{33}}\,
-e^{\frac{19\,\lambda^2\,\left( 4\,m_{N}^2 - s \right) }{33}}
\right]
\nn\\
&& + 
  \frac{432 }{121\,\lambda^2}
     \left[\,  (\chi_{2} + \chi_{3})  ^2 
+ (\chi_{6} +\chi_{7})^2 \right]
\left[e^{\frac{4\,\lambda^2\,\left( 4\,m_{N}^2 - s \right) }{33}}\,
-
e^{\frac{16\,\lambda^2\,\left( 4\,m_{N}^2 - s \right) }{33}}\,
\right]
\nn\\
&&
 + 
  \frac{9  \sqrt{3} }{4\,\lambda^2}
     \left(  \chi_{1} +  \chi_{5} \right) \,\left(  \chi_{4} +  \chi_{8} \right) 
\left[
e^{\frac{\lambda^2\,\left( 4\,m_{N}^2 - s \right) }{12}}
-
e^{\frac{5\,\lambda^2\,\left( 4\,m_{N}^2 - s \right) }{12}}
\right]
\nn\\
&&
 + 
  \frac{27}{{\sqrt{11}}\,\lambda^2}
     \left(  \chi_{2} +  \chi_{3} +  \chi_{6} +  \chi_{7} \right) \,
     \left(  \chi_{4} +  \chi_{8} \right)
\left[
e^{\frac{19\,\lambda^2\,\left( 4\,m_{N}^2 - s \right) }{132}}\,
-
e^{\frac{43\,\lambda^2\,\left( 4\,m_{N}^2 - s \right) }{132}}\,
\right]
\nn\\
&&
 - \frac{27\,}{64}
\left( 4\,m_{N}^2 - s \right) \,
     {\left(  \chi_{4} +  \chi_{8} \right) }^2
\,\,e^{\frac{\lambda^2\,\left( 4\,m_{N}^2 - s \right) }{6}}
-2\,
\left( 4\,m_{N}^2 - s \right) \, \chi_{1}\, \chi_{5} 
\,\,e^{\frac{\lambda^2\,\left( 4\,m_{N}^2 - s \right) }{3}}
\nn\\
&&
- \frac{3456 }{1331}\,
\left( 4\,m_{N}^2 - s \right) \,
     \left(  \chi_{2} +  \chi_{3} \right) \,\left(  \chi_{6} +  \chi_{7} \right) 
\,\,
e^{\frac{10\,\lambda^2\,\left( 4\,m_{N}^2 - s \right) }{33}}\,\,\,.
\label{inn}
\eea
The quantities $\chi_{i}$ are obtained from the sum over the quark color-spin-flavor indices
and are a function of total $NN$ spin $S$ and isospin $I$.
In the appendix A their calculation is shown in detail and in table 1 their values are presented
for different nucleon spin and isospin.

\section{Results of the  simulation  }

\begin{figure*}[t]
\epsfxsize=25pc 
\hspace{2cm}
\epsfbox{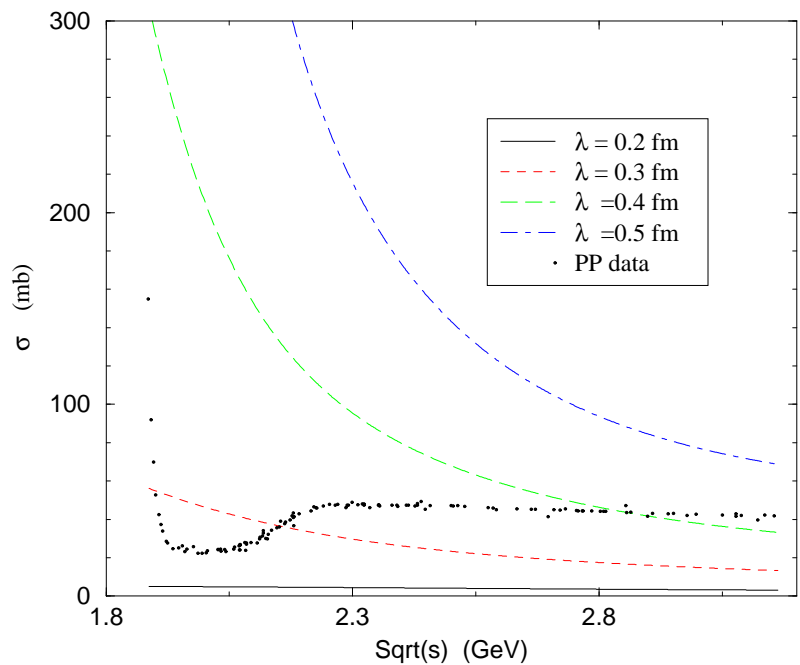}
\caption{(a) The PP cross-section: various values of $\lambda$
compared to the free cross-section labeled as PP data.}
\label{sigma}
\end{figure*}

The $NN$ cross-section is extremely sensitive to the $\lambda$ parameter which is the {\sl rms}
radius of the nucleon. In our calculation the $\Delta$-nucleon mass splitting is set to 0.3 GeV. 
The usual values for $\lambda$ in the quark model range from 0.2 fm to 0.8 fm. 
We have chosen four representative values for $\lambda$ : $0.2,\,0.3,\,0.4,\,0.5$  fm, when combined 
with Eq. (\ref{d-n}) sets the ratio $\alpha_{s}/m_{q}^{2}=0.57,1.97,4.69,9.08$ GeV$^{-2}$.
 The effect of the $\lambda$ variation in the PP cross-section in 
comparison with the free proton-proton cross-section as a function of $\sqrt{s}$ can
be seen in  Fig. \ref{sigma}. 

\begin{table}
\begin{center}
\begin{tabular}{r|c|c|c|c|c}\hline\hline
& $\chi_{1}$ 
&$\chi_{2}$ 
&$\chi_{3}$ 
&$\chi_{4}$ 
& rel. phase $\chi_{5}\ldots\chi_{8}$  
\\\hline
$I=1;$\,$S=1$ 
& $\frac{59}{81}$ 
&  $\frac{17}{81}$ 
& $\frac{17}{81}$
& $\frac{10}{81}$    
&$(-) $                                                  
\\\hline
$S=0$ 
& $\frac{31}{27}$ 
& $\frac{7}{27}$ 
& $\frac{7}{27}$
& $0$
& $(+)$ 
\\\hline\hline
$I=0;$\,$ S=1$ 
& $\frac{19}{27}$ 
& $\frac{7}{27}$ 
& $\frac{7}{27}$ 
& $\frac{2}{27}$
& $(+)$ 
\\\hline
$S=0$
& $-\frac{1}{9}$ 
& $\frac{5}{9}$ 
& $\frac{5}{9}$ 
& $0$
& $(-)$
\\\hline\hline
\end{tabular}
%\label{qbdftab1}
\end{center}
\caption{ $\chi_{i}$ weights as function of total $NN$ spin $S$ and isospin $I$ in 
Eq. (\ref{inn}).  }
\end{table}

The numerical solution of the VUU equation is equivalent to evolving the test particles
by Newtonian mechanics. For the starting configuration, a Fermi-gas ansatz is used and 
all particles of one nucleus are randomly distributed inside a sphere in coordinate and
momentum space, such as
\bea
(\vec{r}_i - \vec{r}_{\rm CM})^2 \leq r^{2}_{0}A^{\frac{2}{3}}
\,\,\,\,\,\,\, ,\,\,\,\,\,\,\,
(\vec{p}_i - \vec{p}_{\rm CM})^2 \leq p^{2}_{F}
\eea
The collision integral in Eq. (\ref{vuu}) is treated in a stochastic way, allowing test 
particles to undergo
collisions with probability proportional to the Pauli-corrected cross-section. A pair of
particles collide if their minimum distance $d$ fulfills the following condition
\bea 
d\leq \sqrt{ \frac{\sigma_{\rm total}}{\pi} } \hspace{1cm}.
\eea

The simulations we present in this paper tested the cross-section variation effects 
on a Niobium-Niobium (Nb+Nb) collision at $E_{lab}=1050$ MeV/nucleon and impact parameter 
$b=3$ fm.  Other nuclei (with symmetric or asymmetric collisions), energies and $b$ have also  
been tested, but our calculation shall be restricted to the Nb+Nb system which exhibits the
qualitative effects we want to demonstrate.
A necessary input for the simulation is the choice of the equation of state. 
We have chosen  a repulsive potential of high compressibility ($k= 380$ MeV) 
the so call  ``hard eos'' with $\alpha=-124$ MeV, $\beta=70.5$ MeV, $\gamma=2$ in Eq. (\ref{skyrme}).
The geometry of the collision is as usual: there are two transverse directions $x$ , $y$ while $z$ is
the beam direction. Thus the  reaction plane is defined as being the $xz$-plane.

\begin{figure}[t]
\epsfxsize=25pc 
\hspace{2cm}
\epsfbox{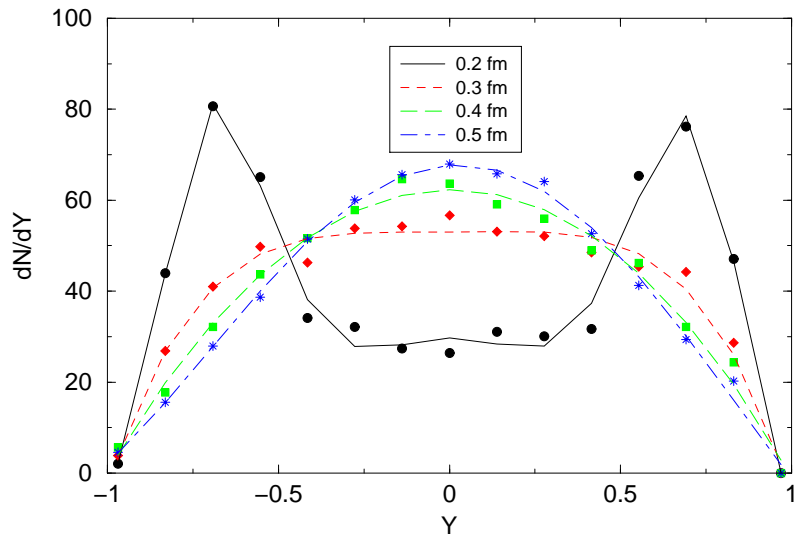}
\caption{Proton distribution as a function of rapidity $Y$ for $E_{lab}=1050$ MeV; $b=3$ fm.}
\label{dndy-y}
\end{figure}

\begin{figure}[t]
\epsfxsize=25pc 
\hspace{2cm}
\epsfbox{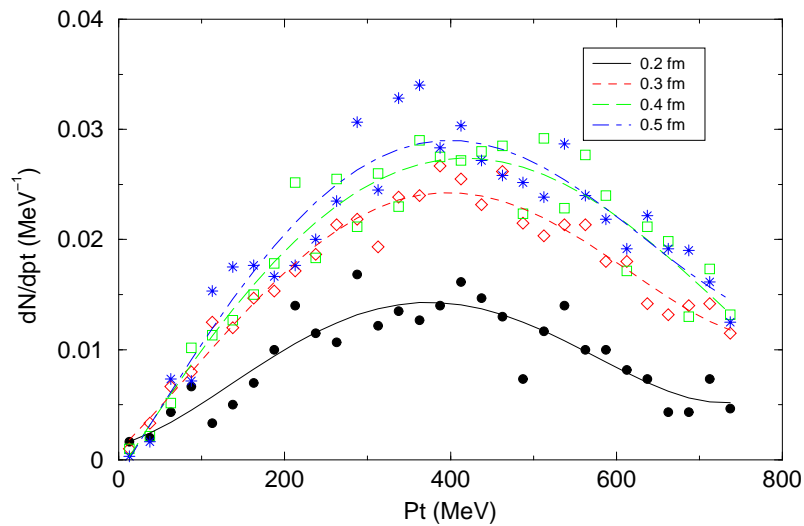}
\caption{Proton distribution as a function of the transverse momentum $p_{t}$ 
for $E_{lab}=1050$ MeV; $b=3$ fm.}
\label{dndpt-pt}
\end{figure}

\begin{figure}[t]
\epsfxsize=25pc 
\hspace{2cm}
\epsfbox{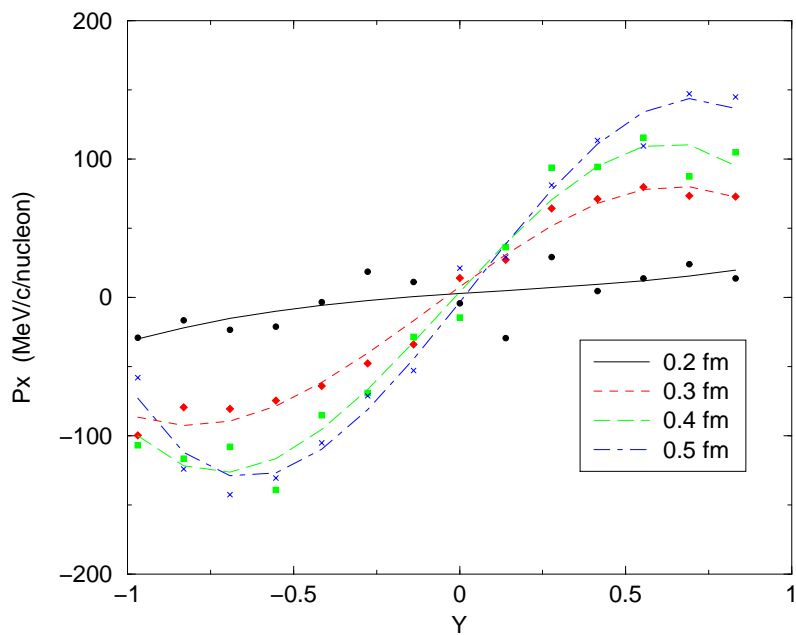}
\caption{In-plane transverse momentum $p_x$ versus rapidity $Y$ for $E_{lab}=1050$ MeV; $b=3$ fm. }
\label{px-y}
\end{figure}

The variation of $\lambda$ affects significantly the  observables as can be seen in Figs.
\ref{dndy-y},\ref{dndpt-pt},\ref{px-y}. In these figures
the curves are average plots where the simulation points are also represented.
The term {\it nuclear stopping power} characterizes the degree of stopping which an incident 
nucleon suffers when it collides with another nucleus and is used in 
the study of  high energy collisions. Due to the strong dependence of the nucleon rapidity distribution
on the $NN$ cross-section a change in $dN/dY$ distribution indicates a creation of a zone higher in
nucleon density.  In Fig. \ref{dndy-y} the proton distribution is shown as a function of rapidity ($Y$),
where a change in the shape of the curve, when $\lambda$ varies from 0.2 fm to 0.5 fm, shows clearly 
the referred effect. The effect of this variation on  the transverse momentum spectra is represented in
Fig. \ref{dndpt-pt}. The transverse momentum distribution ($p_{x}$) as a function of rapidity 
($Y$) is seen in Fig. \ref{px-y}. It becomes evident, from  this figure, that bounce-off also increases 
dramatically as a function of $\lambda$.

\begin{figure}[t]
\epsfxsize=30pc 
\hspace{2cm}
\epsfbox{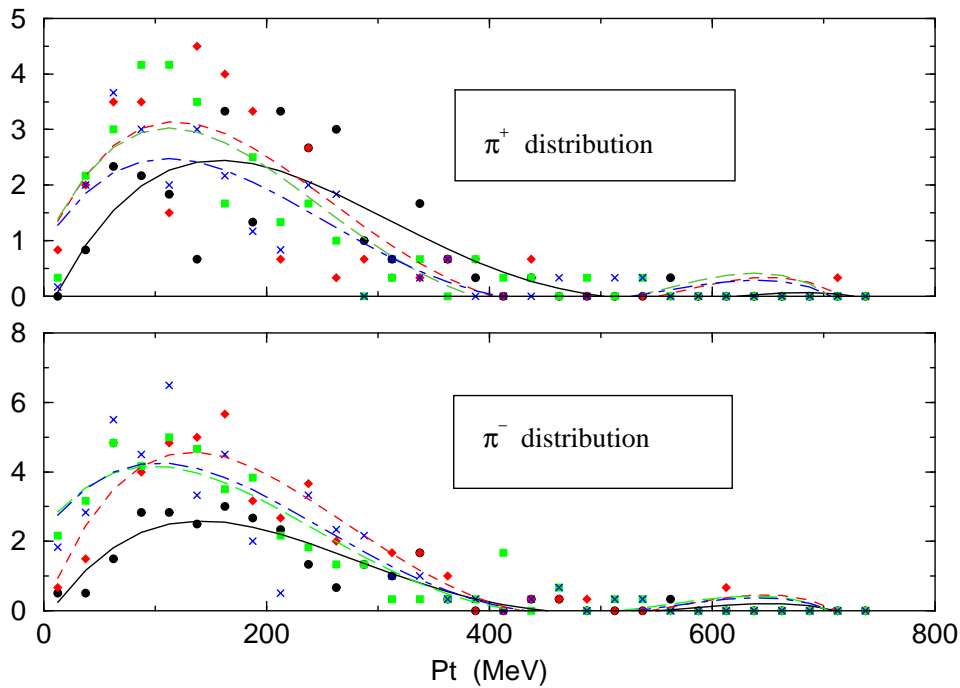}
\caption{The pion distributions (a) for $\pi^{+}$ and for (b) $\pi^{-}$. 
The vertical axis is $dN/dPt$ ($\times 10^{-3}$MeV$^{-1}$).}
\label{pion}
\end{figure}

Although that at current collision energies pion production is present, and incorporated in the 
model through inelastic channels ($NN\rightarrow N\Delta$, $N\Delta\rightarrow NN$, 
$\Delta\rightarrow N\pi$, $N\pi\rightarrow \Delta$) there are no significant modifications in 
the pion distribution as can be seen in Fig. \ref{pion}. 
This can be understood considering that the only modification introduced in this approach is to 
the $\sigma_{NN}$ value.

\section{Conclusions}

In the original VUU model the ions are composed by nucleons which are 
regarded as structureless point particles 
while the equation of state is obtained from the Skyrme potential
$U$ as a function of the density as seen in Eq. (\ref{skyrme}). 
In our model the quark structure of the nucleons are introduced and their effects
are studied in the observables related to the collision. In the very
simple approach we use for the nucleon a single parameter $\lambda$,   
which is the {\it rms} radius,  is treated in the same way as in the naive quark 
models, as a free parameter. In these models it assumes a range of different values in order 
to fit the scattering data. In present study we find that the extended nature of the nucleon,
determined by the finite values of $\lambda$, is actually  reflected in the observables.

In our simulation, as a first approximation,  we use Gaussian wave-function's for the
nucleon which permits an analytical evaluation of the cross-section. A more realistic
choice for $\Psi$ would lead to a multi-dimensional integral in $\sigma_{NN}$, 
which in the context of the present study, represents a more time-consuming and elaborated simulation.
A consequence of using a Gaussian wave-function is an exponential behavior of 
$\sigma_{NN}$ which differs from the free cross-section as can be seen in Fig. 2.
This reflects the simplicity of the model where the repulsive hard-core 
can be considered only in a qualitative aspect.
The complex in-medium effects can be incorporated into the hadron properties in matter 
when a more realistic quark model is considered. For nucleons in-medium an initial 
 wave-function is defined and effective quantaties  $\lambda^{\star}$ and $m_{q}^{\star}$ 
are determined through variational equations \cite{jaime,krein1,krein2}. 
These effects are  beyond the scope of the present calculation. 

Our simulation is also developed in a {\it cold mode}, in which thermal effects 
are not considered. These effects are very important and will be  addressed in a future 
elaboration. An important modification in the interquark potential can 
be obtained from Lattice QCD calculations, for example,  which  indicates that 
strings break in the confining phase at nonzero temperature such as in Ref. \cite{detar}. 

In summary, the introduction of temperature and density effects in the microscopic quark model
is crucial for one to draw definite conculsions. We are aware that the present calculation, in 
this naive approach, is far from this goal, but
has its conceptual importance to the extent that one can relate  many modifications in the
observables to one parameter $\lambda$. 

A possible extension to our current calculation is to consider  Ref. \cite{mb} where a 
meson-baryon interaction is derived in  the context of the Fock-Tani formalism and obtain 
a $\sigma_{\pi N}$ cross-section from quark physics. The influence of the internal quark 
structure to the $\Delta$ ressonance can also  be addressed and is straightforward. 
As stated before, in a more realistic
calculation a density modification, due to the collision, should imply in a change in the 
properties of nucleons and mesons in the medium. In this case an alternative is 
 to simulate the collision  replacing  $\lambda$ and the quark mass $m_{q}$ by effective
 in-medium parameters $\lambda^{*}$ and $m_{q}^{*}$ obtained in a self-consistent
calculation.

Finally, another possible extension to our present work is to address heavy ion collisions at
high energies. One of the most successful models in this regime is the UrQMD  \cite{urqmd}.
From our present approach, general baryon-baryon cross-sections can be calculated,
from \cite{annals} meson-meson cross-sections are obtained and
from \cite{mb} all meson-baryon cross-sections, that appear in UrQMD, can be derived 
incorporating quark interchange.
\section*{Acknowledgments}

The authors would like to thank B. E. J. Bodmann for important suggestions and  H. St\"{o}cker for 
stimulating discussions  held during the  8th International Workshop on Hadron Physics 
2002 at Bento Gon\c{c}alves, Brazil  (April 2002). This research work was supported by 
Funda\c{c}\~ao de Amparo \`a Pesquisa do Rio Grande do Sul (FAPERGS).

\appendix

\section{Evaluation of the $\chi_{i}$ coefficients }

As stated before, details related to the  derivation of the $V_{bb}$ potential in Eq. 
(\ref{hbb}) can be found in Ref. \cite{annals}, so we present here only the final result in 
order to evaluate  the $\chi_{i}$ coefficients
(these coefficients have also been calculated in Ref. \cite{QBD3}):
\bea
V_{bb}(\alpha\beta;\gamma\delta)&=& 
\sum_{i=0}^{4} \,v_{i}(\alpha \beta;\gamma \delta ).
\label{ft93}
\eea
where
\bea
v_{0}(\alpha \beta;\gamma \delta )&=&
9 V_{qq}(\mu \nu; \sig \rho)\,
               \Psi^{\ast\mu\mu_{2}\mu_{3}}_{\alpha} 
                \Psi^{\ast\nu\nu_{2}\nu_{3}}_{\beta} 
                \Psi^{\rho\nu_{2}\nu_{3}}_{\gamma}
                \Psi^{\sig\mu_{2}\mu_{3}}_{\delta}
		\nn\\		
v_{1}(\alpha \beta;\gamma \delta )&=&
-36 V_{qq}(\mu \nu; \sig \rho)\,		
                \Psi^{\ast\mu\mu_{2}\mu_{3}}_{\alpha} 
                \Psi^{\ast\nu\nu_{2}\nu_{3}}_{\beta} 
                \Psi^{\rho\nu_{2}\mu_{3}}_{\gamma}
                \Psi^{\sig\mu_{2}\nu_{3}}_{\delta}		
		\nn\\
v_{2}(\alpha \beta;\gamma \delta )&=&
-9 V_{qq}(\mu \nu; \sig \rho)\,
                \Psi^{\ast\mu\mu_{2}\mu_{3}}_{\alpha} 
                \Psi^{\ast\nu\nu_{2}\nu_{3}}_{\beta}
                \Psi^{\sig\nu_{2}\nu_{3}}_{\gamma}
                \Psi^{\rho\mu_{2}\mu_{3}}_{\delta}
                \nn\\
v_{3}(\alpha \beta;\gamma \delta )&=&
18 V_{qq}(\mu \nu; \sig \rho)\,
                \Psi^{\ast\mu\nu\mu_{3}}_{\alpha} 
                \Psi^{\ast\nu_{1}\nu_{2}\nu_{3}}_{\beta} 
                \Psi^{\rho\nu_{2}\nu_{3}}_{\gamma}
                \Psi^{\nu_{1}\sig\mu_{3}}_{\delta}
		\nn\\
v_{4}(\alpha \beta;\gamma \delta )&=&
-18 V_{qq}(\mu \nu; \sig \rho)\,		
                \Psi^{\ast\mu\mu_{2}\mu_{3}}_{\alpha} 
                \Psi^{\ast\nu_{1}\nu_{2}\nu}_{\beta} 
                \Psi^{\nu_{1}\nu_{2}\mu_{3}}_{\gamma}
                \Psi^{\sig\mu_{2}\rho}_{\delta}
\label{ft94}		
\eea 
Schematically one can represent each  $v_{i}$ in (\ref{ft94}) by
\bea
v_{i}=\chi_{i}\,\,I_{\rm space} (i).
\label{aplic_37.1}
\eea
where $I_{\rm space} (i)$ are the spatial integrals and
\bea
\chi_{i} \equiv {\cal C}(i)\,\,I_{\rm color}(i)\,\,I_{\rm SI}(i) \,\,.
\label{aplic_37.2}
\eea
The factor ${\cal C}$ for each term is obtained directly from (\ref{ft94}) $\times (-1)$
\bea
{\cal C}\,(0,\ldots,4)=(-9,36,9,-18,18).
\label{aplic_38}
\eea
The color factors $I_{\rm color}$ is
\bea
I_{\rm color}(0)&=& \frac{1}{36} \,{\cal F}^{\,a}_{\mu\sigma}{\cal F}^{\,a}_{\nu\rho}\,
             \eps^{\mu\mu_2\mu_3}\eps^{\nu\nu_2\nu_3}
             \eps^{\rho\nu_2\nu_3}\eps^{\sigma\mu_2\mu_3}=0 \nn\\
I_{\rm color}(1)&=& \frac{1}{36} \,{\cal F}^{\,a}_{\mu\sigma}{\cal F}^{\,a}_{\nu\rho}\,
             \eps^{\mu\mu_2\mu_3}\eps^{\nu\nu_2\nu_3}
             \eps^{\rho\nu_2\mu_3}\eps^{\sigma\mu_2\nu_3}=\frac{1}{9} \nn\\
I_{\rm color}(2)&=& \frac{1}{36} \,{\cal F}^{\,a}_{\mu\sigma}{\cal F}^{\,a}_{\nu\rho}\,
             \eps^{\mu\mu_2\mu_3}\eps^{\nu\nu_2\nu_3}
             \eps^{\sigma\nu_2\nu_3}\eps^{\rho\mu_2\mu_3}=\frac{4}{9} \nn\\
I_{\rm color}(3)&=& \frac{1}{36} \,{\cal F}^{\,a}_{\mu\sigma} {\cal F}^{\,a}_{\nu\rho}\,
             \eps^{\mu\nu\mu_3}\eps^{\nu_1\nu_2\nu_3}
             \eps^{\rho\nu_2\nu_3}\eps^{\nu_1\sigma\mu_3}=\frac{2}{9} \nn\\
I_{\rm color}(4)&=& \frac{1}{36} \,{\cal F}^{\,a}_{\mu\sigma}{\cal F}^{\,a}_{\nu\rho}\,
             \eps^{\mu\mu_2\mu_3}\eps^{\nu_1\nu_2\nu}
             \eps^{\nu_1\nu_2\mu_3}\eps^{\sigma\mu_2\rho}=-\frac{2}{9}.
\label{aplic_40}
\eea
where we have used the following property of $SU(N)$ matrices
\bea
\fl
M^{a}_{\mu\sigma}\,M^{a}_{\nu\rho}=
2 \delta_{\mu\rho}\delta_{\nu\sigma}
- f\, 
\delta_{\mu\sigma}\delta_{\nu\rho}
\hspace{.75cm}\mbox{where}\hspace{.5cm}
f=\left\{ 
\begin{array}{ll}
1,              &           \mbox{if $M^{a}=\sigma^{a}$, ($a=1,2,3$)    } \\
 \frac{2}{3},   &           \mbox{if $M^{a}=\lambda^{a}$, ($a=1,\ldots,8$)}
\end{array} \right.   
\label{aplic_39}
\eea
The product  ${\cal C}\,I_{\rm color}$ results
\bea
{\cal C}\,I_{\rm color}(0,\ldots,4)=(0,4,4,-4,-4).
\label{aplic_41}
\eea
The spin-isospin part is 
\bea
I_{\rm SI}(0)&=& \frac{1}{18^2}\, \,
            S^{i}_{s_{\mu}s_{\sigma}}S^{i}_{s_{\nu}s_{\rho}}\,
             \delta_{t_{\mu}t_{\sigma}}\delta_{t_{\nu}t_{\rho}}\,
             \psi^{I_{\mu}I_{\mu_2}I_{\mu_3}}_{\alpha}
	     \psi^{I_{\nu}I_{\nu_2}I_{\nu_3}}_{\beta}
             \psi^{I_{\rho}I_{\nu_2}I_{\nu_3}}_{\gamma}
	     \psi^{I_{\sigma}I_{\mu_2}I_{\mu_3}}_{\delta} \nn\\
I_{\rm SI}(1)&=& \frac{1}{18^2}\, \,
            S^{i}_{s_{\mu}s_{\sigma}}S^{i}_{s_{\nu}s_{\rho}}\,
             \delta_{t_{\mu}t_{\sigma}}\delta_{t_{\nu}t_{\rho}}\,
             \psi^{I_{\mu}I_{\mu_2}I_{\mu_3}}_{\alpha}
	     \psi^{I_{\nu}I_{\nu_2}I_{\nu_3}}_{\beta}
             \psi^{I_{\rho}I_{\nu_2}I_{\mu_3}}_{\gamma}
	     \psi^{I_{\sigma}I_{\mu_2}I_{\nu_3}}_{\delta} \nn\\
I_{\rm SI}(2)&=& \frac{1}{18^2}\, \,
            S^{i}_{s_{\mu}s_{\sigma}}S^{i}_{s_{\nu}s_{\rho}}\,
             \delta_{t_{\mu}t_{\sigma}}\delta_{t_{\nu}t_{\rho}}\,
             \psi^{I_{\mu}I_{\mu_2}I_{\mu_3}}_{\alpha}
	     \psi^{I_{\nu}I_{\nu_2}I_{\nu_3}}_{\beta}
             \psi^{I_{\sigma}I_{\nu_2}I_{\nu_3}}_{\gamma}
	     \psi^{I_{\rho}I_{\mu_2}I_{\mu_3}}_{\delta} \nn\\
I_{\rm SI}(3)&=& \frac{1}{18^2}\, \,
            S^{i}_{s_{\mu}s_{\sigma}}S^{i}_{s_{\nu}s_{\rho}}\,
             \delta_{t_{\mu}t_{\sigma}}\delta_{t_{\nu}t_{\rho}}\,
             \psi^{I_{\mu}I_{\nu}I_{\mu_3}}_{\alpha}
	     \psi^{I_{\nu_1}I_{\nu_2}I_{\nu_3}}_{\beta}
             \psi^{I_{\rho}I_{\nu_2}I_{\nu_3}}_{\gamma}
	     \psi^{I_{\nu_1}I_{\sigma}I_{\mu_3}}_{\delta} \nn\\
I_{\rm SI}(4)&=& \frac{1}{18^2}\, \,
            S^{i}_{s_{\mu}s_{\sigma}}S^{i}_{s_{\nu}s_{\rho}}\,
             \delta_{t_{\mu}t_{\sigma}}\delta_{t_{\nu}t_{\rho}}\,
             \psi^{I_{\mu}I_{\mu_2}I_{\mu_3}}_{\alpha}
	     \psi^{I_{\nu_1}I_{\nu_2}I_{\nu}}_{\beta}
             \psi^{I_{\nu_1}I_{\nu_2}I_{\mu_3}}_{\gamma}
	     \psi^{I_{\sigma}I_{\mu_2}I_{\rho}}_{\delta}.
\label{aplic_42}
\eea
where the spin-isospin index $I_{\mu}\equiv (s_{\mu},t_{\mu} )  $ and $S_{i}=\sigma_{i}/2   $.
\bea
\chi_{0}&=& 0 \nn\\
\chi_{1}&=& \frac{1}{18^2}\, \,
            \sigma^{i}_{s_{\mu}s_{\sigma}}\sigma^{i}_{s_{\nu}s_{\rho}}\,
             \delta_{t_{\mu}t_{\sigma}}\delta_{t_{\nu}t_{\rho}}\,
             \psi^{I_{\mu}I_{\mu_2}I_{\mu_3}}_{\alpha}
	     \psi^{I_{\nu}I_{\nu_2}I_{\nu_3}}_{\beta}
             \psi^{I_{\rho}I_{\nu_2}I_{\mu_3}}_{\gamma}
	     \psi^{I_{\sigma}I_{\mu_2}I_{\nu_3}}_{\delta} \nn\\
\chi_{2}   &=& \frac{1}{18^2}\, \,
            \sigma^{i}_{s_{\mu}s_{\sigma}}\sigma^{i}_{s_{\nu}s_{\rho}}\,
             \delta_{t_{\mu}t_{\sigma}}\delta_{t_{\nu}t_{\rho}}\,
             \psi^{I_{\mu}I_{\mu_2}I_{\mu_3}}_{\alpha}
	     \psi^{I_{\nu}I_{\nu_2}I_{\nu_3}}_{\beta}
             \psi^{I_{\sigma}I_{\nu_2}I_{\nu_3}}_{\gamma}
	     \psi^{I_{\rho}I_{\mu_2}I_{\mu_3}}_{\delta} \nn\\
\chi_{3}&=& -\frac{1}{18^2}\, \,
             \sigma^{i}_{s_{\mu}s_{\sigma}}\sigma^{i}_{s_{\nu}s_{\rho}}\,
             \delta_{t_{\mu}t_{\sigma}}\delta_{t_{\nu}t_{\rho}}\,
             \psi^{I_{\mu}I_{\nu}I_{\mu_3}}_{\alpha}
	     \psi^{I_{\nu_1}I_{\nu_2}I_{\nu_3}}_{\beta}
             \psi^{I_{\rho}I_{\nu_2}I_{\nu_3}}_{\gamma}
	     \psi^{I_{\nu_1}I_{\sigma}I_{\mu_3}}_{\delta} \nn\\
\chi_{4}&=& -\frac{1}{18^2}\, \,
              \sigma^{i}_{s_{\mu}s_{\sigma}}\sigma^{i}_{s_{\nu}s_{\rho}}\,
             \delta_{t_{\mu}t_{\sigma}}\delta_{t_{\nu}t_{\rho}}\,
             \psi^{I_{\mu}I_{\mu_2}I_{\mu_3}}_{\alpha}
	     \psi^{I_{\nu_1}I_{\nu_2}I_{\nu}}_{\beta}
             \psi^{I_{\nu_1}I_{\nu_2}I_{\mu_3}}_{\gamma}
	     \psi^{I_{\sigma}I_{\mu_2}I_{\rho}}_{\delta}.
\label{aplic_43}
\eea
After the contractions of repeated indices in Eq. (\ref{aplic_43}) and 
the correct choices of the nucleon spin and isospin indices ($\alpha$, $\beta$,
$\gamma$, $\delta$) one obtains  the values in table 1.

\end{document}